\documentclass[aps,pre,showpacs,twocolumn,floatfix]{revtex4}

\usepackage{amsmath}
\usepackage{amssymb}
\usepackage{graphicx}
\usepackage{bm}
\usepackage{mathrsfs}

\newcommand{\sech}{\text{sech}}

\begin{document}

\title{Solitons and decoherence in left-handed metamaterials}  

\author{Mattias Marklund$^{1,2}$, Padma K.\ Shukla$^{1,2}$, Lennart Stenflo$^1$ 
  and Gert Brodin$^{1,2}$}
\affiliation{$^1$\,Department of Physics, Ume{\aa} University, SE--901 87 Ume{\aa},
Sweden} 
\affiliation{$^2$\,Centre for Fundamental Physics, Rutherford Appleton Laboratory,
  Chilton Didcot, Oxfordshire, OX11 0QX, UK}

\date{\today}

\begin{abstract}
We present exact electromagnetic solitary pulses that can be experimentally 
obtained within nonlinear left-handed metamaterials. The effect of pulse decoherence
on the modulation instability of partially incoherent electromagnetic waves  is also 
investigated. The results may contribute to a better understanding of nonlinear 
electromagnetic pulse propagation in media with negative index of refraction.
\end{abstract}

\maketitle

The electromagnetic properties of materials exhibiting a negative index of
refraction was investigated already by Mandelstam in 1945 \cite{Mandelstam1,Mandelstam2}.
In addition, the idea of negative permittivity and permeability and the consequences of a negative group
velocity was pointed out by Pafomov \cite{Pafomov} and Agranovich and Ginzburg \cite{Agranovich-Ginzburg}, 
and then discussed by several other authors, e.g. Refs.\ \cite{Veselago,Agranovich1,Agranovich2,Zhang}.
In recent years such materials have been produced \cite{Smith-etal,Shelby-etal}, 
and experimental verification of previous theoretical issues are currently 
being studied with great interest (see, e.g. \cite{Pendry,Ramakrishna} for a review). 
Quite recently, Lazarides and Tsironis \cite{Lazarides-Tsironis} presented a 
model for the propagation of electromagnetic waves in nonlinear metamaterials
having negative index of refraction (left-handed materials).  They also presented 
a study of the possibility of soliton formation within such materials. Nonlinear properties 
of left-handed metamaterials have also been analyzed in Ref. \cite{Zharov}. 

In this Letter, we show that the system presented in Ref.\ \cite{Lazarides-Tsironis}
also allows for dark/bright soliton pairs, exhibiting pulse width broadening due to 
the coupling of the solitons. Moreover, we analyze the statistical properties
of the nonlinear system of equations, in particular the effect of partial
incoherence on the modulational instability of partially incoherent electromagnetic
waves in left-handed metamaterials. The relevance of our investigation to 
laboratory (photonic crystals) and astrophysical settings is discussed. 
 
Left-handed materials simultaneously have negative permittivity $\epsilon_{\text{eff}}$
and permeability $\mu_{\text{eff}}$, and therefore allow for a negative 
refractive index. As demonstrated in Refs.\ \cite{Lazarides-Tsironis} and \cite{Zharov},
it is possible to obtain the expressions $\epsilon_{\text{eff}} = \epsilon + \alpha|\mathbf{E}|^2$
and $\mu_{\text{eff}} = \mu + \beta|\mathbf{H}|^2$, where $\mathbf{E}$ and $\mathbf{H}$ are the 
electric and magnetic fields respectively, for weak fields in left-handed metamaterials.
Here $\epsilon$ and $\mu$ are the linear permittivity and permeability, respectively,
which may take on negative values in metamaterials. We note that if we have focusing media 
($\alpha, \beta > 0$), the nonlinear effect may be used to switch between left-handed and 
right-handed character of the material. Moreover, for defocusing media ($\alpha,\beta < 0$), 
we will have a reinforcement of the left-handed properties of the material. Thus, the nonlinear 
extension of linear metamaterials may have important consequences for future applications. 

Using the expressions for $\epsilon_{\text{eff}}$ and $\mu_{\text{eff}}$, 
Lazarides and Tsironis \cite{Lazarides-Tsironis} derived a system of equations for the
nonlinear propagation of an electromagnetic pulse in left-handed materials, or
\begin{subequations}
\begin{eqnarray} 
  && i\partial_t\psi_a + \partial_x^2\psi_a 
    + (\sigma_a|\psi_a|^2 + \sigma_b|\psi_b|^2)\psi_a = 0 , \\
  && i\partial_t\psi_b + \partial_x^2\psi_b 
    + (\sigma_a|\psi_a|^2 + \sigma_b|\psi_b|^2)\psi_b = 0 , 
\end{eqnarray}
\label{eq:nlse}
\end{subequations}
where the coefficients $\sigma_{a} = \text{sgn}(\alpha\mu)$ and 
$\sigma_b = \text{sgn}(\beta\epsilon)$ can have the values $\pm 1$, 
and the envelope amplitudes $\psi_{a,b}$ (in Ref.\ \cite{Lazarides-Tsironis} 
denoted by $P$ and $Q$) correspond to the electric and
magnetic field components of the electromagnetic pulse, respectively. 
Thus, for focusing media, if 
$\mu,\epsilon > 0$, we have $\sigma_a = \sigma_b = 1$, while 
if $\mu > (<) 0$, $\epsilon < (>) 0$, we have $\sigma_a = +1 (-1)$ and 
$\sigma_b = -1 (+1)$. If both $\mu$ and $\epsilon$ are negative, we
have $\sigma_a = \sigma_b = -1$.
An analysis of the system (\ref{eq:nlse}) was performed in Ref.\ \cite{Lazarides-Tsironis}.
It was found that Eqs.\ (\ref{eq:nlse}) admits stationary solutions of the
form $\psi_i = \psi_{i0}\exp(-i\Delta\omega\,t)$, with $\psi_{i0} > 0$, where the phase 
is determined by $\Delta\omega = -(\sigma_a|\psi_a|^2 + \sigma_b|\psi_b|^2)$. We now perform a 
first order perturbative analysis of Eqs.\ (\ref{eq:nlse}) around the stationary state,
writing $\psi_i = [\Psi_i(x,t) + \psi_{i0}]\exp(-i\Delta\omega\,t)$, where $|\Psi_i| \ll \psi_{i0}$
is a slowly varying envelope of the wave modulation. Linearizing Eqs.\ (\ref{eq:nlse}),
we thus obtain (see, e.g. Ref. \cite{Shukla-Rasmussen})
\begin{equation}\label{eq:lin}
  i\partial_t\Psi_i + \partial_x^2\Psi_i 
  + \psi_{i0}\left[\sigma_a\psi_{a0}(\Psi_a + \Psi_a^*)
  + \sigma_b\psi_{b0}(\Psi_b + \Psi_b^*)\right] = 0 ,
\end{equation}
where the asterisk denotes the complex conjugation. Thus, with $\Psi_i = u_i + iv_i$, 
the real and imaginary parts of Eq.\ (\ref{eq:lin}) give
\begin{subequations}
\begin{eqnarray}
  -\partial_t v_i + \partial_x^2u_i 
  + 2\psi_{i0}(\sigma_a\psi_{a0}u_a + \sigma_b\psi_{b0}u_b) = 0 , \\
  \partial_tu_i + \partial_x^2v_i = 0 ,
\end{eqnarray}
\end{subequations}
respectively. With the ansatz $u_i, v_i \propto \exp(iKx - i\Omega t)$, we obtain
the dispersion relation
\begin{eqnarray}
  && (\Omega^2 - K^4 + 2\sigma_a\psi_{a0}^2K^2 )(\Omega^2 - K^4 
    + 2\sigma_b\psi_{b0}^2K^2)
    \nonumber \\ && \qquad 
     - 4\sigma_a\sigma_b\psi_{a0}^2\psi_{b0}^2K^4 = 0 .
\end{eqnarray}
Thus, we have a modulational instability growth rate $\Gamma = -i\Omega$ 
given by  
\begin{equation}\label{eq:cond}
\Gamma = K\sqrt{2(\sigma_a \psi_{a0}^2 + \sigma_b\psi_{b0}^2) - K^2} . 
\end{equation}
We note that if $\sigma_{a,b} = -1$, the modulational instability is prohibited, reflecting
the well-known fact that one-dimensional dark solitons are modulationally stable.

It was shown in Ref.\ \cite{Lazarides-Tsironis} that Eqs.\ (\ref{eq:nlse}) admit coupled 
bright/bright and dark/dark solitons. In addition, however, it is straightforward to show that 
\begin{equation}
  \psi_a(x,t) =\psi_{a0}\tanh\left(\frac{x - 2Kt}{L}\right)\exp(iKx - i\Omega_at), 
\end{equation}
and
\begin{equation}
  \psi_b(x,t) =\psi_{b0}\,\sech\left(\frac{x - 2Kt}{L}\right) \exp(iKx - i\Omega_bt) , 
\end{equation}
are solutions \cite{Vladimirov-etal} to Eqs.\ (\ref{eq:nlse}), if $L^{-2} 
= (\sigma_b\psi_{b0}^2 - \sigma_a\psi_{a0}^2)/2$,
$\Omega_a = -K^2 - \sigma_a\psi_{a0}^2$, and $\Omega_b = -K^2 
- (\sigma_a\psi_{a0}^2 + \sigma_b\psi_{b0}^2)/2$. These coupled dark/bright solitons are
solutions, which seem to have been overlooked in Ref.\ \cite{Lazarides-Tsironis}. The effects
of the nonlinear coupling between dark and bright solitons, such as the width dependence
on the signs of $\sigma_{a,b}$ and the pulse intensities, should be experimentally verifiable. 

The statistical properties of electromagnetic pulse propagation in 
nonlinear metamaterials can be studied using the Wigner formalism. 
Introducing the Wigner distribution functions (see, e.g.\ \cite{Wigner})
\begin{equation}
  F_i(x,t,k) = \frac{1}{2\pi}\int\,d\xi\,e^{ik\xi}\psi_i^*(x + \xi/2,t)\psi_i(x - \xi/2,t) ,
\end{equation}
where $i = a, b$. We then obtain the coupled kinetic (or the Wigner-Moyal) equations
\cite{Besieris-Tappert, Mendonca, Anderson-etal3}
\begin{subequations}
\begin{equation}
  \partial_tF_a + 2k\partial_xF_a 
  + 2(\sigma_a|\psi_a|^2 + \sigma_b|\psi_b|^2)\sin\left(
   \tfrac{1}{2}\stackrel{\leftarrow}{\partial_x}\stackrel{\rightarrow}{\partial_k}\right)F_a = 0 , 
\end{equation}
and
\begin{equation}
  \partial_tF_b + 2k\partial_xF_b 
  + 2(\sigma_a|\psi_a|^2 + \sigma_b|\psi_b|^2)\sin\left(
   \tfrac{1}{2}\stackrel{\leftarrow}{\partial_x}\stackrel{\rightarrow}{\partial_k}\right)F_b = 0 ,  
\end{equation}
from Eqs.\ (\ref{eq:nlse}). Since
\begin{equation}
  |\psi_i(x,t)|^2 = \int\,dk\,F_i(x,t,k) ,
\end{equation}
\label{eq:system}
\end{subequations}
Eqs.\ (\ref{eq:system}) form a closed set of equations for partially incoherent 
electromagnetic pulses in left-handed metamaterials. 

Next, we consider the first order perturbation of Eqs.\ (\ref{eq:system}). We 
let $F_i(x,t,k) = F_{i0}(k) + F_{i1}(x,t,k)\exp(iKx - i\Omega t)$, where
$F_{i1} \ll F_0$. Linearizing Eqs.\ (\ref{eq:system}) we obtain the 
dispersion relation
\begin{eqnarray}
  1 &=& -\frac{1}{2K}\int\,dk\,\Bigg\{ 
  \frac{\sigma_a[F_{a0}(k + K/2) - F_{a0}(k - K/2)]}{k - \Omega/2K}
  \nonumber \\ &&\quad
    + \frac{\sigma_b[F_{b0}(k + K/2) - F_{b0}(k - K/2)]}{k - \Omega/2K}
    \Bigg\}.
    \label{eq:dispersion}
\end{eqnarray}

A monochromatic pulse can be analyzed by using the ansatz $F_{i0} = I_{i0}\delta(k - k_{i0})$, where $I_{i0} = |\psi_{i0}|^2$ is the intensity 
of the field $\psi_{i0}$, in Eq.\ (\ref{eq:dispersion}). On the other
hand, if there is a stochastically varying phase in the background envelope fields 
$\psi_{i0}$, the corresponding Wigner function is given by the Lorentzian distribution
\begin{equation}\label{eq:lorentz}
  F_{i0}(k) = \frac{I_i}{\pi}\frac{k_{iT}}{(k - k_{i0})^2 + k_{iT}^2} ,
\end{equation}
where $k_{iT}$ is the width of the distribution around $k_{i0}$. With the 
distribution (\ref{eq:lorentz}), the dispersion relation (\ref{eq:dispersion}) 
takes the form 
\begin{eqnarray}
  1 &=& -\frac{2\sigma_aK^2I_a}{[\Omega - 2(k_{a0} - i k_{aT})K]^2 - K^4}
  \nonumber \\ &&
        -\frac{2\sigma_bK^2I_b}{[\Omega - 2(k_{b0} - i k_{bT})K]^2 - K^4} .
  \label{eq:dispersion2}
\end{eqnarray}
In many cases, it is reasonable to assume that the statistical spread is equal 
and centered around the same value for the two envelopes $\psi_{a,b0}$. i.e. 
$k_{a0,T} = k_{b0,T} = k_{0,T}$. Then Eq.\ 
(\ref{eq:dispersion2}) yields
\begin{equation}
   [\Omega - 2(k_{0} - i k_{T})K]^2 - \left[ K^2 - 2(\sigma_aI_a + \sigma_bI_b) 
   \right]K^2  
   = 0.
  \label{eq:dispersion3}
\end{equation}
Letting $\Omega =2k_0 K + i \Gamma$ in Eq. (\ref{eq:dispersion3}), we obtain the growth 
rate $\Gamma$ for $\sigma_i I_a + \sigma_b I_b > K^2/2$. We have 
\begin{equation}
   \Gamma = K\sqrt{2(\sigma_aI_a + \sigma_bI_b) - K^2\,}\, -  2k_T K,
  \label{eq:growth}
\end{equation}
which shows the damping effect of the decoherence spread $k_T$.
Moreover, as noted above, there is no growth in the case of $\sigma_{a,b} = -1$.
Equation (\ref{eq:growth}) also shows the influence of a focusing metamaterial. 
When $\mu < 1$, but $\epsilon >1$, we have $\sigma_a = -1$ and 
$\sigma_b = +1$, removing the modulational instability altogether 
when $I_a \geq I_b$.

To summarize, we have presented nonlinear solutions for electromagnetic fields
in materials with negative index of refraction. In addition, we have 
also discussed the modulational instability of partially incoherent electromagnetic
waves in such materials by employing the Wigner-Moyal equation. Choosing the
equilibrium Wigner function in the form of a Lorentzian distribution, corresponding to
a random phase, we have deduced a nonlinear dispersion, which exhibits an oscillatory 
modulational instability whose growth rate is given by Eq.\ (\ref{eq:growth}). The present
results should be useful for understanding the nonlinear propagation of 
finite amplitude electromagnetic pulses in left-handed metamaterials in
laboratory and astrophysical (e.g.\ rotating black holes \cite{Mackay-etal}) environments.
For example, multiple black holes and other massive objects in outer space
can make the light beams bend in unexpected and unpredictable ways. The
unexpected light bending can be attributed to the nonlinear propagation of  
intense electromagnetic pulses in a medium with negative index of refraction.

\end{document}